\documentclass[3p,times,procedia]{elsarticle}
\usepackage{nupha_ecrc}
\usepackage{amsmath}
\usepackage{amssymb}

\volume{00}

\firstpage{1}

\journalname{Nuclear Physics A}

\runauth{}

\jid{nupha}

\jnltitlelogo{Nuclear Physics A}

\usepackage{amssymb}

\usepackage[figuresright]{rotating}

\begin{document}

\begin{frontmatter}



\dochead{XXVIIIth International Conference on Ultrarelativistic Nucleus-Nucleus Collisions\\ (Quark Matter 2019)}

\title{The  crossover line in the $(T, \mu)$-phase diagram of QCD}


%
\author[a,b]{Jana N. Guenther\footnote{speaker: Jana.Guenther@t-online.de}}
\author[b]{Szabolcs Bors\'anyi}
\author[b,c,d,e]{Zoltan Fodor}
\author[b]{Ruben Kara}
\author[d]{Sandor D. Katz}
\author[b]{Paolo Parotto}
\author[b]{Attila P\'asztor}
\author[f]{Claudia Ratti}
\author[b,c]{Kalman~K.~Szab\'o}

\address[a]{Department of Physics, University of Regensburg, Universitätsstraße 31, 93053 Regensburg, Germany}
\address[b]{Department of Physics, University of Wuppertal, Gaussstra\ss e 20, 42119 Wuppertal, Germany}
\address[c]{J\"ulich Supercomputing Centre, Forschungszentrum J\"ulich, 52425
J\"ulich, Germany}
\address[d]{Institute for Theoretical Physics, E\"otv\"os University, H-1117 Budapest, Hungary}
\address[e]{Physics Department, UCSD, San Diego, CA 92093, USA}
\address[f]{Department of Physics, University of Houston, Houston, TX 77204, USA}
\begin{abstract}
An efficient way to study the QCD phase diagram at small finite density
is to extrapolate thermodynamical observables from imaginary chemical potential.
The phase diagram features a crossover line starting from the
transition temperature already determined at zero chemical potential.
In this work we focus on the Taylor expansion of this line up to $\mu^4$ contributions. We present
the continuum extrapolation of the crossover temperature based on different observables at several lattice spacings. 
\end{abstract}


\end{frontmatter}


\section{Introduction}
\label{}
\begin{figure}
\begin{minipage}{0.55\textwidth}
  \includegraphics[width=\textwidth]{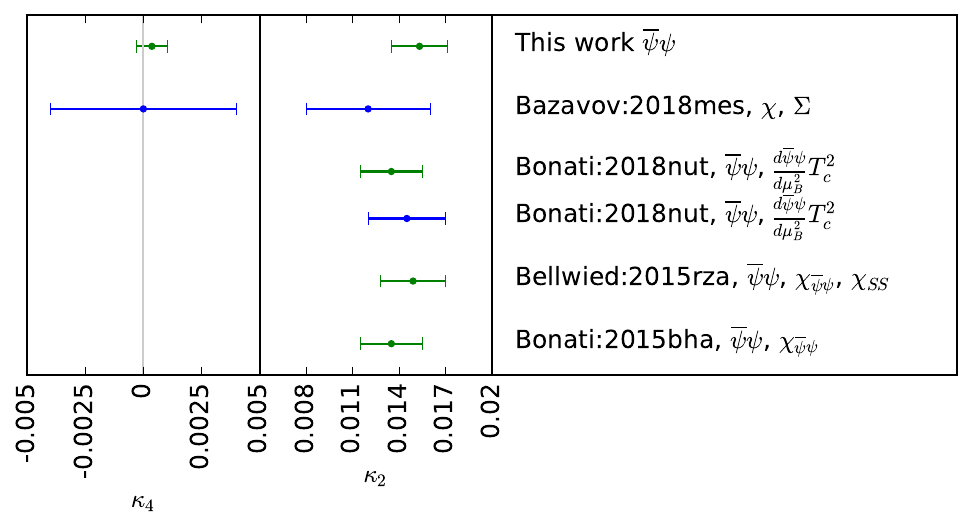}
\caption{A comparison of recent results for $\kappa_2$ and $\kappa_4$ as defined in eq.~(\ref{eq:kappa}). \cite{Bazavov:2018mes} and the lower point of \cite{Bonati:2018nut} use the Taylor method, while the result from this work, the upper point of \cite{Bonati:2018nut}, \cite{Bellwied:2015rza} and \cite{Bonati:2015bha} used lattice data at imaginary chemical potential.\label{fig:kappa}}
\end{minipage}
\hfill
\begin{minipage}{0.39\textwidth}
 \includegraphics[width=\textwidth]{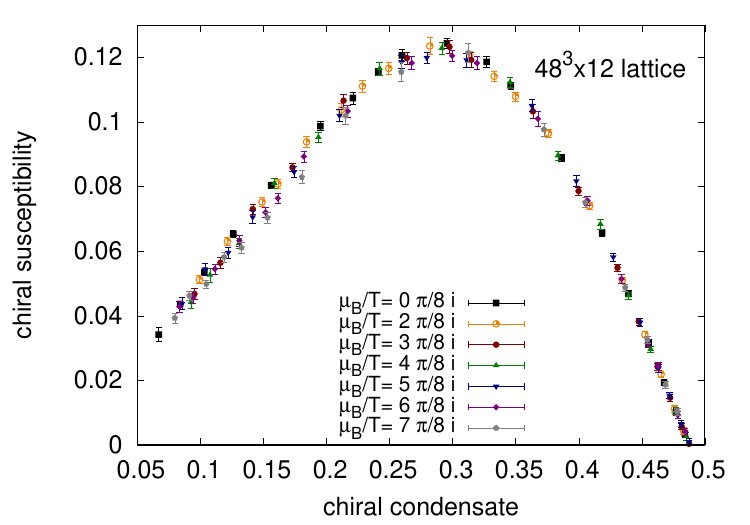}\\
 \caption{The chiral susceptibility as a function of the chiral condensate on an $48^3\times12$ lattice for different imaginary chemical potentials. Most of the dependence of $\mu_B$ is removed and the data collapses to one curve.\label{fig:pbp}}
\end{minipage}
\end{figure}

An important question in the study of QCD is the investigation of the $(T, \mu_B)$-phase diagram. Due to the infamous sign problem direct lattice simulations are restricted to vanishing or imaginary baryon chemical potential. However, since it is established that at $\mu_B = 0$ the transition is an analytic crossover \cite{Aoki:2006we, Bhattacharya:2014ara}, this opens the possibility to gain knowledge about the $(T, \mu_B)$-plane by analytical continuation. In this proceedings we address the extrapolation of the crossover line from imaginary to real  chemical potential as done in \cite{Bonati:2018nut, Bellwied:2015rza,Cea:2015cya,Bonati:2015bha}. Another approach to gain the same results is the determination of the Taylor coefficients at vanishing chemical potential (\cite{Bazavov:2018mes,Bonati:2018nut,Kaczmarek:2011zz,Endrodi:2011gv}). With both methods it is possible to determine the coefficients $\kappa_2$ and $\kappa_4$ which describe the transition line as
\begin{equation}
 \frac{T_c(\mu_B)}{T_c(0)} = 1 - \kappa_2 \left(\frac{\mu_B}{T_c} \right)^2  - \kappa_4 \left(\frac{\mu_B}{T_c} \right)^4   + \mathcal O(\mu_B^6).\label{eq:kappa}
\end{equation}
A comparison of recent results is done in fig.~\ref{fig:kappa}. Both methods agree well with each other. In this proceedings we summarize the updated results for $\kappa_2$ and $\kappa_4$ recently published in \cite{Borsanyi:2020fev}. 

\section{Analysis}
\begin{figure}
 \begin{center}
  \includegraphics[width=0.64\textwidth]{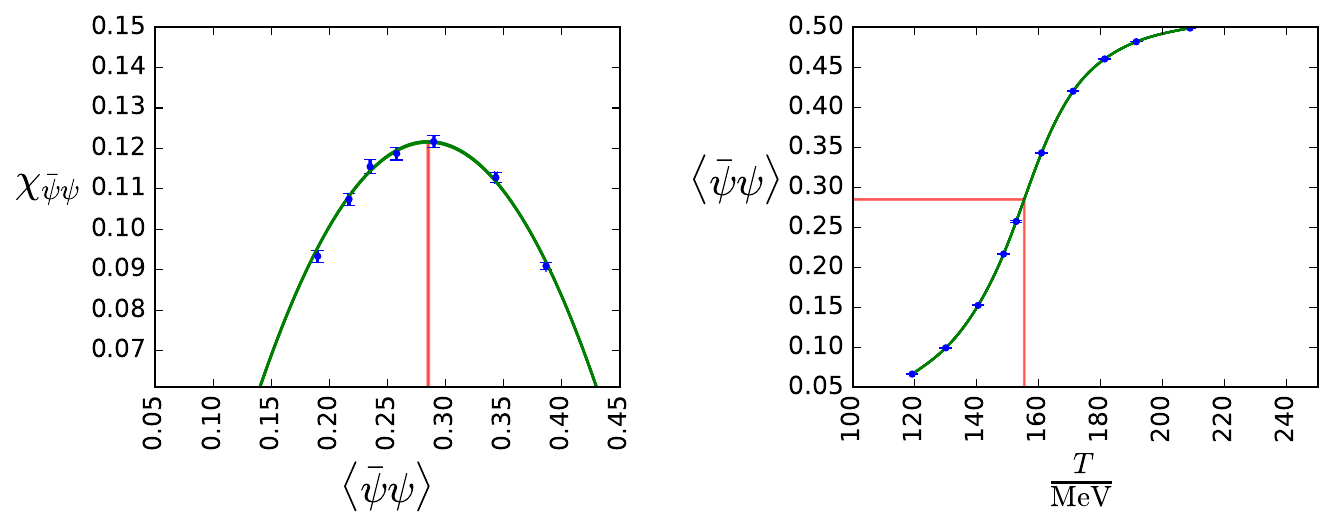} \hfill
  \includegraphics[width=0.34\textwidth]{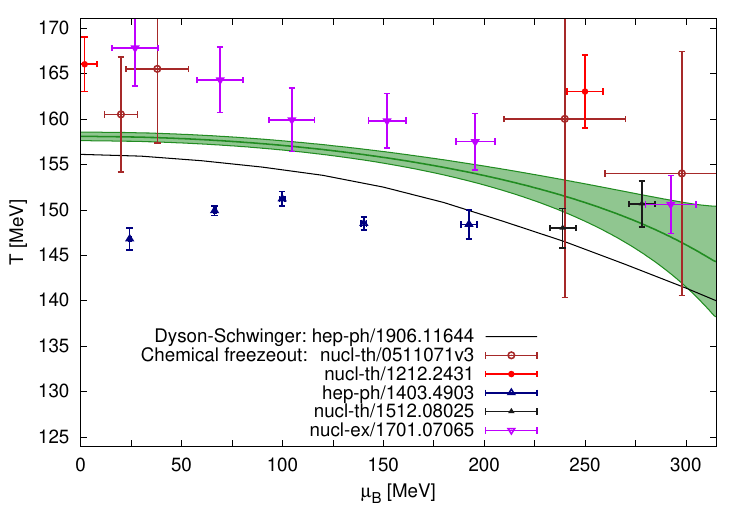}
 \end{center}
\caption{Left and middle: Illustration of the determination of the crossover temperature on an $40^3\times10$ lattice at $\mu_B=0$. First the position of the peak in the chiral susceptibility as a function of the chiral condensate is determined. In a second step a spline is used to translate the value of the chiral condensate to a temperature. Right: The extrapolation of the cross over line to finite chemical potential. The  band was obtained the from Lattice simulations described in this proceedings. The line is the result from \cite{Isserstedt:2019pgx} calculated with Dyson-Schwinger equations. Please note, that the absolute value at  $\mu_B=0$ is set by a lattice computation. Therefor a comparison should only be done for the curvature. The points from \cite{Andronic:2005yp, Becattini:2012xb, Alba:2014eba,Vovchenko:2015idt, Adamczyk:2017iwn} illustrate the freeze-out regime in heavy ion collisions.\label{fig:illustration}}
\end{figure}

In this work we are investigating the chiral condensate $ \langle\bar\psi\psi\rangle$ and the chiral susceptibility $\chi_{\bar\psi\psi}$. The bare quantities are the derivatives of the partition function $Z$ with respect to the quark masses:
\begin{align}
  \langle\bar\psi\psi\rangle^\text{bare} = \frac{T}{V} \frac{\partial \ln Z}{\partial m_q}, \ \
  &\chi_{\bar\psi\psi}^\text{bare} = \frac{T}{V} \frac{\partial^2 \ln Z}{\partial m_q ^2}.
\end{align}
For renormalization the finite and zero temperature values have to be subtracted, yielding the renormalized quantities:
\begin{align}
  \langle\bar\psi\psi\rangle &= \left(\langle\bar\psi\psi\rangle^\text{bare}(0, \beta) - \langle\bar\psi\psi\rangle^\text{bare}(T, \beta) \right) \frac{m_l}{ f_\pi^4}, \ \
  \chi_{\bar\psi\psi} = \left(\chi_{\bar\psi\psi}^\text{bare}(T, \beta) - \chi_{\bar\psi\psi}^\text{bare}(0, \beta) \right) \frac{m_l^2}{ f_\pi^4}.
\end{align}
Due to the analytic nature of the transition the definition of the crossover temperature is ambiguous. In this analysis we capitalize on an observation which can be made in fig.~\ref{fig:pbp}: If one considers the chiral susceptibility as a function of the chiral condensate most of the dependence of $\mu_B$ is removed and the data collapses for different values of imaginary chemical potential. This allows us to fit $\chi_{\bar\psi\psi}$ for fixed $N_t$ but various $\mu_B$ with the ansatz:
\begin{equation}
  \chi_{\bar\psi\psi}(\langle\bar\psi\psi\rangle) = \sum_{i=0}^n \alpha_i\left( 1+ \beta_{i}\left( \frac{\mu_B}{T} \right)^2 \right) \langle\bar\psi\psi\rangle^i,
 \end{equation}
 $n\in\{2,3,4\}$ for appropriate fit ranges. 
This fit removes most of the $\mu_B$ dependence and allows for a precise determination of the transition value of $\langle\bar\psi\psi\rangle$ at the peak. In a next step this has to be translated into temperature. To determine the temperature from the the $\langle\bar\psi\psi\rangle$ value we use a spline. This procedure is illustrated in fig.~\ref{fig:illustration}. 

After determining the transition temperature for various imaginary chemical potentials and lattice spacings, we perform a combined extrapolation in $\frac{\mu_B^2}{T^2}$ and to the continuum, which uses a fully correlated fit. From a mock analysis we learned that, to reliably extract $\kappa_2$ and $\kappa_4$ with our current accuracy on the transition temperatures we need to use at least three free parameters to describe the $\mu_B$ dependence in our fit. Therefore we use these two different fit functions for the $\frac{\mu_B^2}{T^2}$ direction:
\begin{align}
\frac{T_c(\mu_B)}{T_c(0)}&=1+a\left( \frac{\mu_B}{T} \right)^2+b\left( \frac{\mu_B}{T} \right)^4+c\left( \frac{\mu_B}{T} \right)^6,\label{eq:extrap1} \ \
\frac{T_c(\mu_B)}{T_c(0)}= \frac{1}{1+a\left( \frac{\mu_B}{T} \right)^2+b\left( \frac{\mu_B}{T} \right)^4+c\left( \frac{\mu_B}{T} \right)^6}.
\end{align}
To estimate the error on our results we do 768 different analyses, varying fit functions and ranges as well as the scale sitting. We determine a systematic error by looking at the width of the distribution of the different results. However we only keep values where the $Q$-value of the fits is lager then 0.1. The statistical error is estimated by the Jackknife method and both errors are added in quadrature. This error measures how precisely we can determine the crossover temperature (or its Taylor coefficients) for one definition of the transition. It does not measure the width of the transition region, which has to be investigated in an additional analysis.

\section{Results}
To extrapolate the crossover line to real chemical potential we use the two fit functions from eq.~\ref{eq:extrap1}. Our result is shown by the band in the right side of fig.~\ref{fig:illustration}. We also compare our extrapolation to several other recently published results. The line is the result from \cite{Isserstedt:2019pgx} calculated with Dyson-Schwinger equations. One should note, that the absolute value at  $\mu_B=0$ is set by a lattice computation. Therefor an comparison should only be done for the curvature, which agrees well with both of the bands, obtained by lattice calculations. The points from \cite{Andronic:2005yp, Becattini:2012xb, Alba:2014eba,Vovchenko:2015idt, Adamczyk:2017iwn} illustrate the freeze-out regime in heavy ion collisions. This does not have to be directly on the curve as the transition region is wider than the error band for $T_c$. However they are expected to be in the same temperature range, which they are.

Our results for $\kappa_2$ and $\kappa_4$ as defined in eq.~\ref{eq:kappa}, are 
\begin{equation}
 \kappa_2 = 0.0153 \pm 0.0018, \ \ \kappa_4 = 0.00032 \pm 0.00067.
\end{equation}
They are shown in fig.~\ref{fig:kappa}. The value for $\kappa_2$ agrees well with previous determinations, both from the Taylor and the imaginary $\mu_B$ method. For $\kappa_4$ we do the second determination of this quantity and reduce the error significantly, confirming the expectation that $\kappa_4\ll\kappa_2$.

\clearpage

\clearpage
\section{Acknowledgements}
This project was funded by the
DFG grant SFB/TR55. This work was supported by the
Hungarian National Research, Development and Innovation Office, NKFIH grants KKP126769 and K113034.
An award of computer time was provided by the INCITE
program. The authors gratefully acknowledge the Gauss
Centre for Supercomputing e.V. (www.gauss-centre.eu)
for funding this project by providing computing time on
the GCS Supercomputer JURECA/Booster at Jülich Supercomputing Centre (JSC), on HAZELHEN at HLRS,
Stuttgart as well as on SUPERMUC-NG at LRZ, Munich. This material is based upon work supported by
the National Science Foundation under grants no. PHY-
1654219 and by the U.S. Department of Energy, Office of
Science, Office of Nuclear Physics, within the framework
of the Beam Energy Scan Topical (BEST) Collaboration.
C.R. also acknowledges the support from the Center of
Advanced Computing and Data Systems at the University of Houston. A.P. is supported by the János Bolyai
Research Scholarship of the Hungarian Academy of Sciences and by the ÚNKP-19-4 New National Excellence
Program of the Ministry for Innovation and Technology.





\bibliographystyle{elsarticle-num}
\bibliography{QM}







\end{document}